\documentclass[12pt]{article}
\usepackage[colorlinks,linkcolor=blue,citecolor=blue,urlcolor=blue,bookmarks,bookmarksnumbered]{hyperref}
\usepackage{amsmath,amssymb,amsfonts,mathrsfs}
\usepackage[paper=letterpaper,margin=1.0in]{geometry}
\usepackage{graphicx,xcolor}
\RequirePackage{tikz}
 \usetikzlibrary{arrows}
 \def\TikZ#1{\begin{tikzpicture}#1\end{tikzpicture}}
\usepackage{cite}

\newcommand{\be}{\begin{equation}}
\newcommand{\ee}{\end{equation}}
\newcommand{\bea}{\begin{eqnarray}}
\newcommand{\eea}{\end{eqnarray}}
\newcommand{\ba}{\begin{array}}
\newcommand{\ea}{\end{array}}
\newcommand{\ben}{\begin{enumerate}}
\newcommand{\een}{\end{enumerate}}
\newcommand{\bi}{\begin{itemize}}
\newcommand{\ei}{\end{itemize}}
\newcommand{\bc}{\begin{center}}
\newcommand{\ec}{\end{center}}
\newcommand{\bfig}{\begin{figure}}
\newcommand{\efig}{\end{figure}}
\newcommand{\bq}{\begin{quotation}}
\newcommand{\eq}{\end{quotation}}
\newcommand{\bt}{\begin{table}}
\newcommand{\et}{\end{table}}
\newcommand{\btab}{\begin{tabular}}
\newcommand{\etab}{\end{tabular}}
\newcommand{\bs}{\begin{slide}}
\newcommand{\es}{\end{slide}}

\newcommand{\IC}{\mathbb{C}}
\newcommand{\IP}{\mathbb{P}} 
\newcommand{\IR}{\mathbb{R}}
\def\ZZ{\mathbb{Z}}

\newcommand{\vev}[1]{\langle #1 \rangle}

\newcommand{\rd}{\mathrm{d}}

\def\s{\sigma}

\def\s{\sigma}

\let\a=\alpha
\def\tA{\tilde{A}}
\let\ba=\overline
\def\tB{\tilde{B}}

\let\d=\delta

\def\rd{{\rm d}}
\def\define{\buildrel{\rm def}\over=}
\let\f=\phi
\let\F=\Phi
\def\e{\epsilon}
\let\F=\Phi

\def\inv#1{{\textstyle{1\over#1}}}

\let\L=\Lambda
\def\Lb{\Lambda_b}
\let\q=\theta

\let\p=\pi

\let\t=\tau
\let\vd=\partial
\def\vev#1{\left\langle#1\right\rangle}

\let\w=\omega

\def\frc#1#2{{\textstyle{#1\over#2}}}
\def\RR{\relax\leavevmode
       \ifmmode\mathchoice
       {\hbox{\sf R\kern-.4em R}}
       {\hbox{\sf R\kern-.4em R}}
       {\lower.9pt\hbox{\scriptsize\sf R\kern-.36em R}}
       {\lower1.2pt\hbox{\tiny\sf R\kern-.36em R}}
       \else{\sf R\kern-.4em R}\fi}

\def\resetby#1#2{\@addtoreset{#2}{#1}}
\def\seceq{\@addtoreset{equation}{section}
              \def\theequation{\thesection.\arabic{equation}}}

\def\Label#1{\label{#1}%
                \smash{\hbox to0pt{\raise1ex\hbox{\tiny[#1]}\hss}}}
\def\noLabels{\let\Label=\label}
\def\Eq#1{Eq.~(\ref{#1})}

\def\pd#1#2{\frac{\partial#1}{\partial#2}}
\DeclareMathOperator{\SL}{\textrm{SL}}

\DeclareMathOperator{\sech}{\textrm{sech}}
\DeclareMathOperator{\sgn}{\textrm{sgn}}

\let\sss=\scriptscriptstyle
\def\BHM{`axilaton'}

\def\cS{\mathcal{S}}
\def\sW{\mathscr{W}}
\def\sX{\mathscr{X}}
\def\sY{\mathscr{Y}}
\def\sZ{\mathscr{Z}}
\def\Dj{\rule[.75ex]{4pt}{.6pt}\kern-4ptD}
\def\dj{d\kern-4pt\rule[1.15ex]{4pt}{.5pt}}

\parskip\medskipamount
\begin{document}

\noindent
\begin{minipage}{\hsize}
\begin{center}
\vspace{7mm}
\centerline{\LARGE\bfseries On Stringy de~Sitter Spacetimes}
\vspace{7mm}

\centerline{{\bf
	Per Berglund${}^{1}$\footnote{\tt per.berglund@unh.edu},
	Tristan H{\"u}bsch${}^{2}$\footnote{\tt thubsch@howard.edu}
and
	Djordje Mini{\'c}${}^{3}$\footnote{\tt dminic@vt.edu}
}}
\vspace{3mm}

{\footnotesize\it
${}^1$Department of Physics and Astronomy, University of New Hampshire,
 Durham, NH 03824, U.S.A. \\
${}^2$Department of Physics and Astronomy, Howard University, Washington,
 D.C., 20059, U.S.A. \\
${}^3$Department  of Physics, Virginia Tech, Blacksburg, VA 24061, U.S.A.\\
}
\end{center}

\begin{abstract}\noindent
We reexamine a family of models with a 3+1-dimensional de~Sitter spacetime obtained in the standard tree-level low-energy limit of string theory with a non-trivial anisotropic axion-dilaton background. While such limiting approximations are encouraging but incomplete, 
 our analysis reveals a host of novel features, and shows these models to interpolate between standard and well understood supersymmetric string theory solutions. Finally, we conjecture that this de~Sitter spacetime naturally arises by including more of the stringy degrees of freedom, such as a recently advanced variant of the phase-space formalism, as well as the analytic continuation of a complex two-dimensional Fano variety arising as a small resolution in a Calabi-Yau 5-fold.
\end{abstract}

\end{minipage}

\section{Introduction, Results and Synopsis} 
\label{s:IRS}
Whether or not asymptotically de~Sitter spacetime can exist as a solution of string theory has been one of the fundamental conundrums in string theory ever since the dramatic discovery of 
dark energy in the late 1990s~\cite{Riess:1998cb, Perlmutter:1998np}.
This question is still considered open~\cite{Danielsson:2018ztv,Cicoli:2018kdo}, and the interest in this hard and fundamental issue has been reignited recently~\cite{Obied:2018sgi, Agrawal:2018own}.

Within the standard low-energy limit of string theory, we focus on the effective action for Einstein's gravity (we concentrate on the observed 4-dimensional case) in the familiar format (we adopt the ``mostly positive'' metric signature throughout):
\begin{equation}
 S_{\text{eff}} = \int d^4 x \sqrt{g} \Big({-}\frac{1}{8 \pi G} \L + \frac{1}{16 \pi G} R
 +a R_{\mu \nu} R^{\mu \nu} +b R^2 + cR_{\mu \nu \rho \sigma} R^{\mu \nu \rho \sigma} +\dots\Big).
 \label{e:Seff}
\end{equation}
The coefficients $a,b,c$ as well as the omitted metric/curvature terms are completely determined by the renormalization of the underlying worldsheet theory~\cite{rF79b,rF79a}, and the usual formulation of (target) spacetime in string theory, identified with the vev's of certain ``coordinate'' quantum fields in the underlying worldsheet field theory~\cite{rBZ-StrTh,rGSW1,rJPS}. Also omitted from~\eqref{e:Seff} are ``matter'' terms, the relevant of which are discussed below.

In particular, we will focus on a class of models arising from type IIB/F-theory, the so called {\em\/\BHM\ models\,}\footnote{The admittedly playful name is a reminder that the deformation family of the spacetime varying string vacua 
of Ref.~\cite{rBHM1, rBHM2, rBHM3, rBHM4, rBHM5, rBHM6} are driven by the background values of the axion-dilaton system and their axial $\SL(2;\ZZ)$ monodromy in a transversal 2-dimensional plane $\sY^2$, around the non-compact spacetime, $\sW$. The succinct name also saves us from repeated circumlocutions that perforce include this string of references.}~\cite{rBHM1,rBHM2,rBHM3,rBHM4,rBHM5,rBHM6}:
 ({\small\bf1})~which evade the oft-mentioned no-go theorem\footnote{For other examples that evade the no-go theorem, see~\cite{Gibbons:2001wy}.}~\cite{Gibbons:1984kp,
Maldacena:2000mw,rBBS,Gran:2018ijr},
 ({\small\bf2})~interpolate between two classes of standard supersymmetric string theory solutions~\cite{rFTh,rS-ForFld} and~\cite{Einhorn:2000ct}, and a third, possibly novel class,
 ({\small\bf3})~form a {\em\/discretuum\/} owing to their stringy $\SL(2;\ZZ)$ monodromy, and 
 ({\small\bf4})~require $g_s\,{\sim}\,O(1)$, with an effective incorporation of S-duality. This resonates with some recent assessments~\cite{Andriot:2018wzk}, and some features of the recent efforts~\cite{Heckman:2018mxl,Heckman:2019dsj}; it reminds of the ``T-fold'' constructions~\cite{Dabholkar:2002sy,Flournoy:2004vn,rH-nGeoCY,rHIS-nGeoCY}, and qualifies the standard low-energy effective theory limit description as encouraging but {\em\/incomplete\/}: It indicates a need to include more of the stringy degrees of freedom, as also advocated recently in the phase-space approach~\cite{Freidel:2013zga, Freidel:2014qna, Freidel:2015pka, Freidel:2015uug, Freidel:2016pls, Freidel:2017xsi, Freidel:2017wst, Freidel:2017nhg, Freidel:2018apz}, and earlier, in the double field theory approach~\cite{Tseytlin:1990nb,Tseytlin:1990va,Siegel:1993th,Siegel:1993xq,Hull:2009mi}.

Section~\ref{s:BHM} summarizes the \BHM\ models as an iterative deformation of Minkowski supersymmetric string (and F-)theory compactifications into de~Sitter solutions
with cosmologically broken supersymmetry. For simplicity, these models are driven by the axion-dilaton background configuration in the F-theory formulation of type-IIB string theory~\cite{rFTh}, with the above-listed key results discussed and analyzed in Section~\ref{s:Features}.

 In Section~\ref{s:Other}, we discuss analogous constructions driven by the dynamics of compactification moduli, which then resemble the T-folds~\cite{Dabholkar:2002sy,Flournoy:2004vn,rH-nGeoCY,rHIS-nGeoCY} wherein (T-dual) mirror-symmetry is involved in patching local charts. The \BHM\ models involve S-duality in a similar vein, so that these two construction types combine easily and define a very general class of models. Also, the Minkowski\,$\to$\,de~Sitter metric deformation
  correlates with the desingularization of the total spacetime in \BHM\ models, implying (if tentatively) that 3+1-dimensional de~Sitter spacetime occurs {\em\/generically\/} within string theory, via so-called {\em\/exceptional\/} subspaces~\cite{rHitch}.
 
 Foremost, and in view of the incompleteness discussed in Section~\ref{s:Features}, these characteristics jointly imply that a more accurate description of de~Sitter spacetime in string theory can only be achieved beyond the standard low-energy effective field theory limit.

\section{The Iterative Deformation Models}
\label{s:BHM}
We now turn to reviewing the class of theories of interest, with the goal of setting the stage for an improved understanding of the resulting de~Sitter spacetimes, to which we will turn in the latter sections of the paper. 
The \BHM\ models were developed~\cite{rBHM1,rBHM2,rBHM3,rBHM4,rBHM5,rBHM6} as an iterated deformation (and partial decompactification) of standard string compactifications, 
focusing for simplicity on the special case driven by the dynamics of the axion-dilaton system in the F-theory description of type-IIB string theory~\cite{rFTh}. Herein, we provide a roadmap to this iterative deformation and specify the notation, setting the stage for deriving and discussing the above-itemized key results in Section~\ref{s:Features}, and generalizations in Section~\ref{s:Other}.

\paragraph{A Roadmap:}
The \BHM\ deformation family of models is constructed by starting with an
  F-theoretic type-IIB string theory spacetime, $\sW^{3,1}\times\sY^6(\times T^2)$, where the complex structure of the zero-size ``hidden'' $T^2$ fiber of F-theory is identified with the axion-dilaton $\t\define\a\,{+}\,ie^{-\F}$ modulus~\cite{rFTh}\footnote{As usual, only the complex structure of the $T^2$ is relevant, with the volume having been shrunk to zero.}.
\begin{enumerate}
 \vspace{-5pt}\itemsep=-3pt\vspace*{-1mm}\setcounter{enumi}{0}
 \item\label{i:Holo}
  Deform this \`a la stringy cosmic strings~\cite{rGSVY,rCYCY} into\footnote{Here, we borrow the group-theoretic symbol ``$\rtimes$'' to denote that $\sW^{3,1}$ varies (is fibered) over $\sY^2$.}
  $\sW^{3,1}\rtimes\sY^2\times\sY^4$, where:
  \begin{enumerate}\itemsep=-3pt\vspace*{-1mm}
   \item
    $\t$ and the observable spacetime $\sW^{3,1}$ (via warped metric) vary over $\sY^2$,
   \item\label{i:infty}
    $\sY^2\to S^1\times\sZ$, with the polar parametrization $re^{i\q}=\ell e^{z+i\q}$,
   \item
    $\sY^4=\text{K3}$ or $T^4$ preserves supersymmetry\footnote{It is straightforward to also {\em\/fiber\/} $\sY^4$ over  $\sY^2$, thus relating the \BHM\ models to the virtually ubiquitous K3- and elliptic fibration models; see Section~\ref{s:Other}.}.
  \vspace*{-5pt}
  \end{enumerate}
  While $\t$ (and the ``hidden'' $T^2$) is holomorphic over $\sY^2\,{\approx}\,\IC^1=(\IP^1{\,\smallsetminus}\,\{\infty\})$, $\sW^{3,1}_{(z_i,\q_i)}$ are cosmic 3-branes at special isolated points $(z_i,\q_i)\in\sY^2$~\cite{rGSVY,rCYCY}.

 \item\label{i:nHolo}
  Deform $\t$ to vary {\em\/non-holomorphically,} only over $S^1\,{\subset}\,\sY^2$, while the metric varies only over $\sZ\,{\subset}\,\sY^2$, ---and turns complex beyond $z_0$, which locates the circular naked singularity. As the proper distance to both ends of $z\,{\in}\,\sZ\,{=}\,({-}\sgn(z_0){\cdot}\infty,z_0)$ is infinite, $\sZ\approx\IR^1$ always: $\sY^2\,{\approx}\,\IC^1$ has been punctured into $S^1\times\sZ$ and the cosmic branes of step~\ref{i:Holo} in the above roadmap have effectively coalesced to $z\to z_0$ and $-\sgn(z_0){\cdot}\infty$.

 \item\label{i:patch}
  Cross-patching the two distinct solutions, $\sZ\,{=}\,({-}\sgn(z_0){\cdot}\infty,z_0)$, at $z\,{=}\,0$ into two annuli/cylinders: one with the naked singularity at the ends of $\sZ\approx\IR^1$ and one without, but both with extra, $\d(z)$-localized matter required by matching conditions~\cite{rBHM2}.

 \item\label{i:Lb}
  Deforming the $\sW^{3,1}_{\!z=0}$ metric to de~Sitter ($\Lb>0$) removes the spacetime curvature singularities at $z_0\in\sZ$ of the $\Lb\to0$ Minkowski limit.
\end{enumerate}
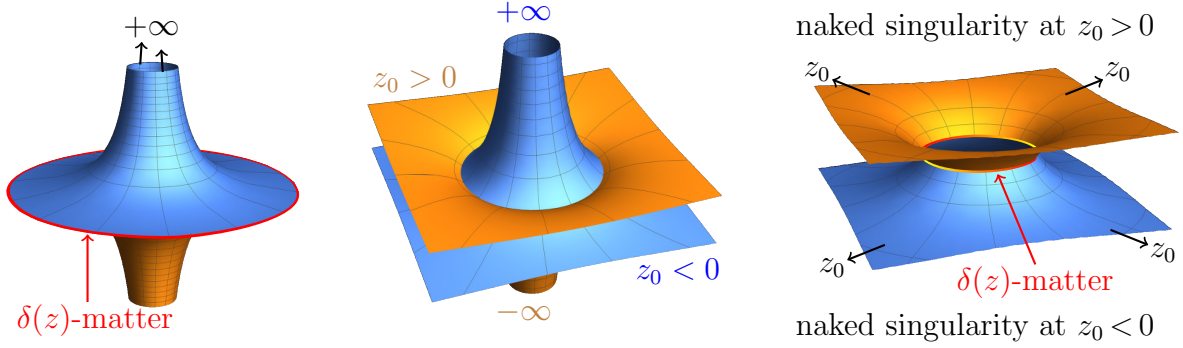
\begin{figure}[htb]
 \begin{center}
  \begin{picture}(160,35)(0,4)
   \put(0,0){\includegraphics[width=50mm]{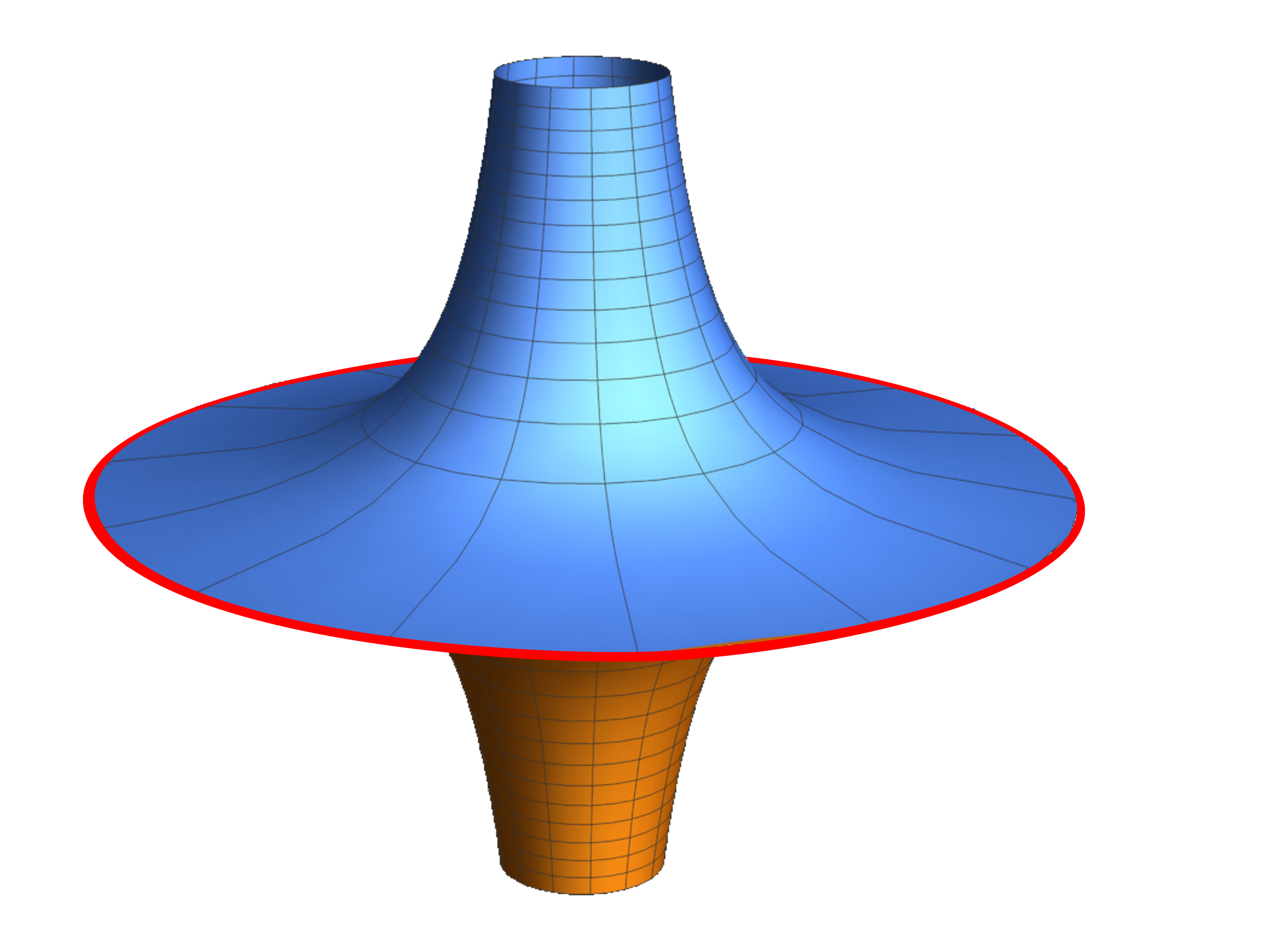}}
    \put(0,0){\TikZ{[scale=2]\path[use as bounding box](0,0);
               \path[red](.75,0)node{$\d(z)$-matter};
               \draw[thick,red,->](.7,.12)--++(0,.45);
               \draw[thick,->](1.05,1.7)--++(.01,.15);
               \draw[thick,->](1.2,1.65)--++(-.01,.15);
                \path(1.125,1.95)node{$+\infty$};
                    }}
   \put(50,0){\includegraphics[width=50mm]{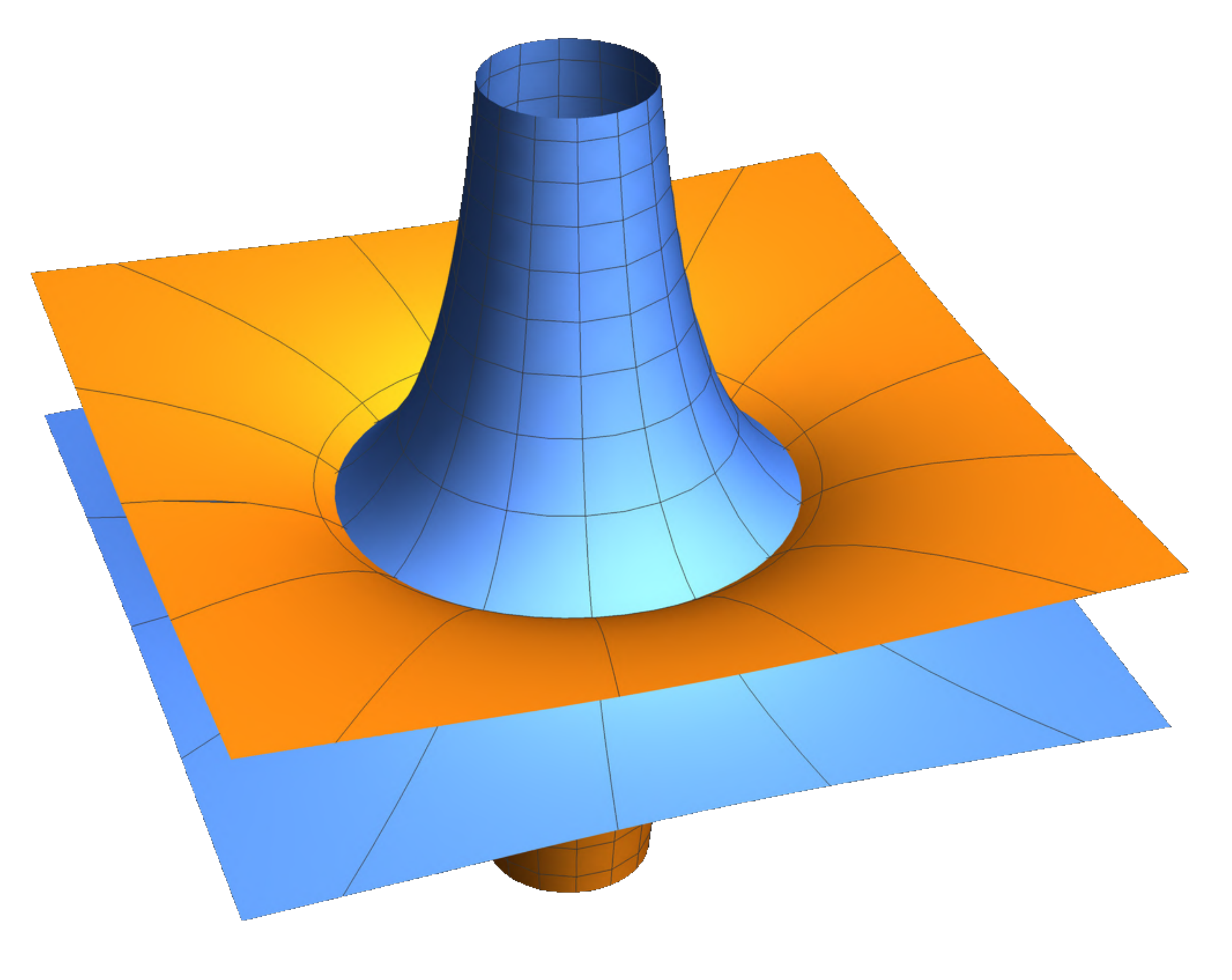}}
    \put(50,0){\TikZ{\path[use as bounding box](0,0);
                \path[brown](0.75,3.2)node{$z_0>0$};
                 \path[brown](2.2,.1)node{$-\infty$};
                \path[blue](4.25,.65)node{$z_0<0$};
                 \path[blue](2.2,4.1)node{$+\infty$};
                     }}
   \put(110,5){\includegraphics[width=50mm]{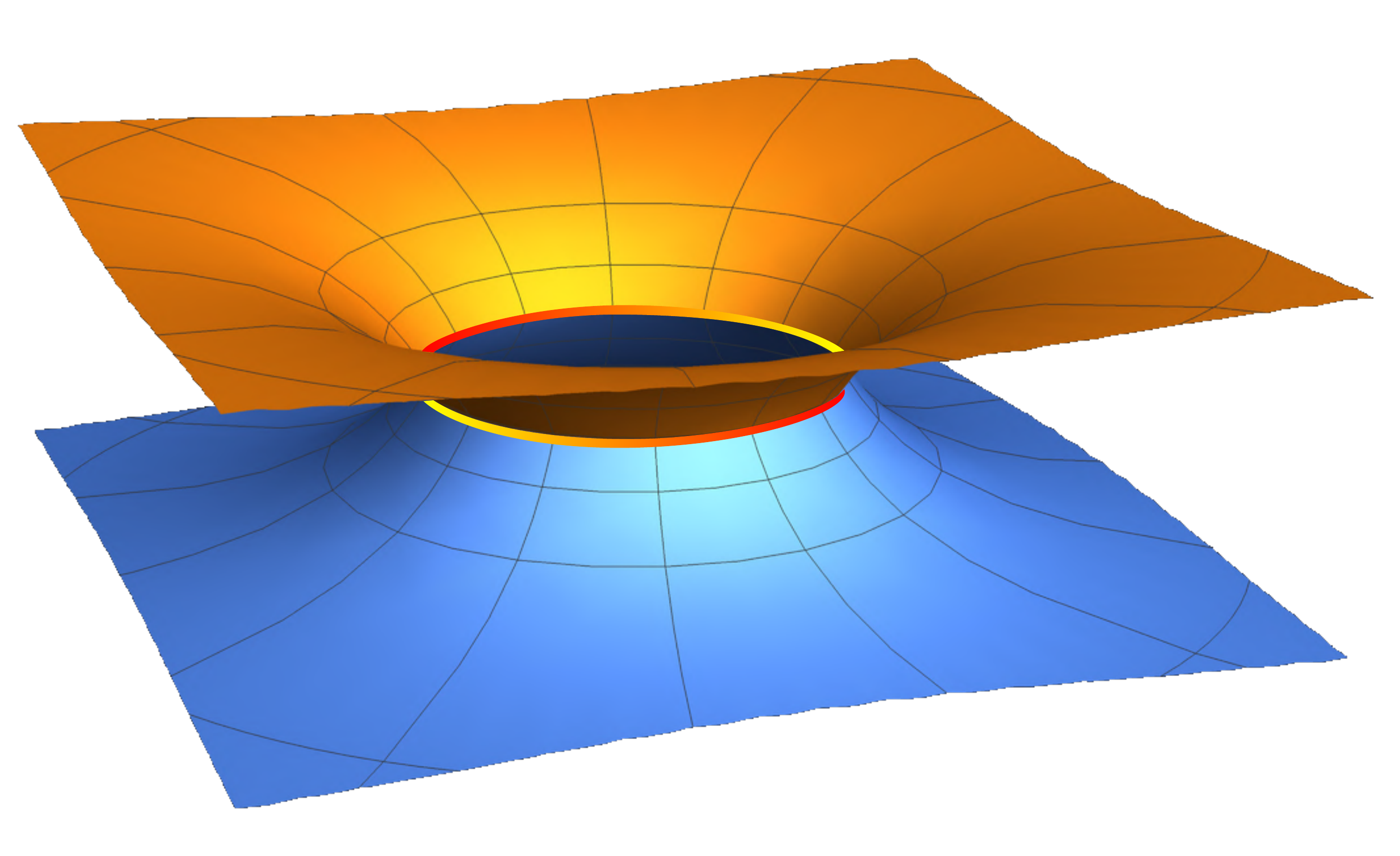}}
    \put(110,5){\TikZ{[scale=2]\path[use as bounding box](0,0);
                 \path[red](1.5,0)node{$\d(z)$-matter};
                 \draw[thick,red,->](1.5,.12)--++(-.25,.6);
                 \draw[thick,->](1.7,1.25)--++(.25,.1);
                  \path(2.05,1.4)node{$z_0$};
                 \draw[thick,->](.4,1.25)--++(-.25,.1);
                  \path(.05,1.4)node{$z_0$};
                 \draw[thick,->](2,.35)--++(.25,-.1);
                  \path(2.35,.2)node{$z_0$};
                 \draw[thick,->](.5,.25)--++(-.25,-.1);
                  \path(.15,.1)node{$z_0$};
                 \path(1.1,1.7)node{naked singularity at $z_0\,{>}\,0$};
                 \path(1.1,-.3)node{naked singularity at $z_0\,{<}\,0$};
                      }}
  \end{picture}
 \end{center}
 \caption{Plotting (vertically) the proper distance against (radially) the circumference in the horizontal plane; the two simple solutions (center), patched solutions (left and right)}
 \label{f:charts}
\end{figure}
As noted in steps~\ref{i:nHolo} and~\ref{i:patch} of the above roadmap, proper distance to both ends of $z\in({-}\sgn(z_0)\infty,z_0)$ is infinite, as is the proper circumference at $z_0$; see Figure~\ref{f:charts}. Therefore, all versions of $\sY^2$ in the \BHM\ configurations are diffeomorphic to annuli/cylinders, which are non-compact Calabi-Yau 1-folds.

\paragraph{Warping:}
While $D=6$ is the phenomenologically relevant case, where $\sW^{3,1}_{\!z=0}$ is $D{-}2=3{+}1$-dimensional, working in a general $D$-dimensional background highlights the inherent $D$-dependence. Having fixed $\sY^4=\text{K3}$ or $T^4$, we omit this factor until Section~\ref{s:Other}.

The codimension-2 solution $\sW^{3,1}\rtimes(S^1\times\sZ)$ of the final step~\ref{i:Lb} in the above roadmap, has a positive cosmological constant, $\Lb$, along $\sW^{3,1}$, and the warped metric is:
\begin{subequations}
 \label{e:metric}
\begin{alignat}9
           \rd s^2 &= A^2(z)\, \bar g_{ab}\,\rd x^a \rd x^b
                     + \ell^2 B^2(z)\,(\rd z^2 + \rd\q^2)~, \label{e:Metric} \\[1mm]
           \bar g_{ab}\,\rd x^a \rd x^b &= - \rd x_0^2 +
                e^{2\sqrt{\Lb}\,x_0}\,(\rd x_1^2 + \ldots + \rd x_{D-3}^2)~, \label{e:gbar}
\end{alignat}
\end{subequations}
where $z=\log(r/\ell)\in\sZ$~\cite{rBHM5}.

\paragraph{The Modulus:}
The two explicit solutions for $\t$ satisfying the above separation of variables~\cite{rBHM1,rBHM2}:
\begin{subequations}
 \label{e:moduli}
\begin{alignat}9
  \t_I(\q) &=b_0+i\,g_s^{-1}\,e^{\w(\q-\q_0)}, \label{e:tI}\\
 \t_{I\!I}(\q) &=\big(b_0\pm g_s^{-1}\tanh[\w(\q{-}\q_0)]\big)
               \pm i\,g_s^{-1}\sech[\w(\q{-}\q_0)] \label{e:tII}
\end{alignat}
\end{subequations}
are anisotropic and aperiodic over $|\q|\leq\p$, but exhibit a non-trivial $\SL(2;\ZZ)$ monodromy for {\em\/specific choices\/} in the effective parameter space $(b_0,\w,g_s)$; the $\q_0$-dependence in~\eqref{e:moduli} may be absorbed by suitably redefining $g_s$~\cite{rBHM1}. With the Teichm{\"u}ller metric ${\cal G}_{\t \bar{\t}}=-1/(\t{-}\bar\t)^2$, the ${\cal G}_{\t\bar\t}|\partial\t|^2$ addition to the action~\eqref{e:Seff} is $\SL(2;\ZZ)$-invariant.

\paragraph{The Metric:}
With the metric~\eqref{e:Metric} axially symmetric while $\t$ is independent of the radial distance from the cosmic brane~\eqref{e:moduli}, the Einstein equation simplifies to:
\begin{equation}
         R_{\mu\nu} ={\cal G}_{\t\bar{\t}}\,\vd_{\mu}\t\vd_{\nu}\bar\t
         \define\widetilde{T}_{\mu\nu}=\hbox{diag}[0,{\cdots},0,\inv4\w^2\ell^{-2}]~,\qquad
         \text{with}~~{\cal G}_{\t \bar{\t}}=-\frac1{(\t{-}\bar\t)^2}.
\label{e:EinStein}
\end{equation}
Thereby, although the total spacetime $\sW^{3,1}\,{\rtimes}\,\sY^2\,{\times}\,\sY^4(\times T^2)$ in fact does admit a Ricci-flat metric, the background \BHM\ configuration~\eqref{e:moduli} drives the spacetime metric to deform away from the Ricci-flat choice.
 This deviation from Ricci-flatness is clearly characterized~\eqref{e:EinStein}
 by the anisotropy $\w\in\IR$ (so $\w^2>0$) that induces supersymmetry breaking~\cite{rBHM1}, and
 by the characteristic (transversal) length scale $\ell$ in $\sY^2$.

For $\w\neq0$ ($\t\neq\hbox{\it const.}$), one has a
perturbative, analytic solution\footnote{This solution is of the same
form
as that discussed by Gregory~\cite{Gregory:1996dd,Gregory:1999gv} for the $U(1)$ vortex solution.}:
\begin{subequations}
 \label{e:newAB}
\begin{alignat}9
    A(z) &= Z(z) \Big(1- \frac{\w^2 z_0^2(D{-}3)}{24(D{-}1)(D{-}2)} Z(z)^2 + O(\w^4)\Big)~,
\label{e:newA}\\
   B(z) &= \frac{1}{\ell z_0\sqrt{\Lb}}\Big(1 -\frac{\w^2z_0^2}{8(D{-}1)} Z(z)^2 + O(\w^4)\Big)~, \label{e:newB}
\end{alignat}
\end{subequations}
where $Z(z)\define1{-}z/z_0$ for $z_0>0$.
Spacetime is asymptotically flat approaching the horizon $z\,{=}\,z_0$ (at infinite proper distance)~\cite{rBHM5}, in agreement with the behavior of Rindler space~\cite{Kaloper:1999sm}.

In stark contrast, the $\Lb\to0$ limiting Minkowski solution is very different~\cite{rBHM1,rBHM2}:
\begin{equation}
           \tA(z) = Z(z)^{\frac{1}{(D-2)}}~,\qquad
           \tB(z) = Z(z)^{-\frac{(D-3)}{2(D-2)}}
                   e^{-\frac12(Z(x){-}1)[\xi(Z(z){+}1){+}2]}~,
\label{e:oldAB}
\end{equation}
and exhibits a naked singularity at $z\,{=}\,z_0$; this generalizes the
 $\xi\,{=}\,-1$ case studied in Ref.~\cite{rBHM5}; $\xi$ counts cosmic branes in units of $\pm\frac{1}{12}$~\cite{rGSVY}. Most importantly: the $\Lb\,{\to}\,0$ limit does not change the transverse geometry, only the metric along the cosmic brane.

\paragraph{Phenomenology:}
Whereas $\Lb>0$ (i.e., $\w^2>0$) desingularizes the metric~\eqref{e:metric} with~\eqref{e:newAB},
\Eq{e:oldAB} is still a good approximation to \Eq{e:newAB} away from $z_0$\footnote{This was first shown by Gregory~\cite{Gregory:1996dd,Gregory:1999gv} and later realized in the present context in Ref.~\cite{rBHM5}.}.
In particular, by comparing \Eq{e:newAB} with \Eq{e:oldAB} close
to the core one can show that~\cite{rBHM1,rBHM5}:
\begin{subequations}
 \label{e:link}
\begin{alignat}9
  z_0 &=-\frac{h}{ h'}\Big|_{z=0}~,\qquad\qquad 
  \xi=\Big(\frac{h''}{2 h'} - \frac{\omega^2 h}{ 8 h'}\Big)\Big|_{z=0},\qquad
  \w_{\sss\text{GCB}}^2\define8\xi/z_0,\label{e:a0-xi}\\
  \ell &= \Lb^{-1/2}\sqrt{ \frac{h'' h^{-\frac{(D-4)}{(D-2)}}}{ (D-2)(D-3)}}\bigg|_{z=0}~,
\qquad h(z)\define A(z)^{D-2}=(1{-}z/z_0)^{D-2}.
\label{e:ell-lambda}
\end{alignat}
\end{subequations}
That is, given a smooth solution defined by~\eqref{e:newAB} and parameterized in terms of $(z_0,\omega,\Lb)$,  this solution close to $z=0$ can be interpreted as a 
cosmic brane solution with parameters $(z_0,\xi,\ell)$ determined by the $(z_0,\omega,\Lb)$ through Eqs.~\eqref{e:link}. 
 Alternatively, we can solve for $\Lb$,
\begin{equation}
\Lb=\frac{\Big(\omega^2 - \omega^2_{\sss\text{GCB}} A^2|_{z=0}\Big)}{4 \ell^{2} (D-2)(D-3)}
\define\frac{\Delta \omega^2}{4 \ell^{2} (D-2)(D-3)}~.
\label{e:dS=omega}
\end{equation}
The anisotropy $\omega$ determines the stress tensor~\eqref{e:EinStein} for the de~Sitter ($\Lb\,{>}\,0$) solution, and asymptotes to the Minkowski ($\Lb\,{\to}\,0$)
cosmic brane limit $\omega_{\sss\text{GCB}}$ at $z\,{\to}\,0$. 
The cosmological constant is thus directly related to the non-trivial
variation of the modulus $\t$---and thereby the string coupling constant---as a function of $\q$!
This gives a very non-trivial relation between the stringy moduli,
and hence string theory itself, and a positive $\Lb$.
Furthermore, $\Lb>0$ is equivalent\footnote{That $\Lb$ {\em\/must\/} be positive follows directly from the defining equation for the warp factor $A(z)$~\cite{rBHM5}.} to $\omega^2 > \omega_{\sss\text{GCB}}^2$, so that $\omega^2 =0$ also implies that $\omega^2_{\sss\text{GCB}}=0$. The latter being a necessary condition for restoring supersymmetry establishes the important relation between supersymmetry breaking and a positive cosmological constant.

Finally, the Newton constant, $G^{(D-2)}_N=M_{D-2}^{-(D-4)}$, in
$D{-}2$ dimensions and the zero-mode wave function normalization,
$\vev{\psi_0|\psi_0}$, are~\cite{rBHM5}:
\begin{equation}
      G^{(D-2)}_N=
     M_{D}^{-(D-2)} \vev{\psi_0|\psi_0}^{-1}~,\qquad\mbox{and}\qquad
\vev{\psi_0|\psi_0}\sim \frac{\pi}{D{-}3}\frac{\ell}{\sqrt{\Lb}}~.
\label{GN}
\end{equation}
The volume of the transversal space, $V_\perp=\vev{\psi_0|\psi_0}$,
is large~\cite{rBHM5} and drives the large $M_{D-2}/M_D$ hierarchy.
This then implies the following relation,
\begin{equation}
\L_{D-2}\sim \Big(\frac{\pi}{D{-}3}\Big)^2
M_{D-2}^{~D-2}~(\ell\,M_{D-2})^2~\Big(\frac{M_D}{M_{D-2}}\Big)^{2D-4}~,
   \label{e:energydensity}
\end{equation}
where $\L_{D-2}=\Lb/G^{(D-2)}_N$ is the energy density in $D{-}2$ dimensions.

\paragraph{4 Dimensions:}
Focusing on the phenomenologically relevant case of $D=6$, recall that $\ell$ is the characteristic (transverse) size of the cosmic brane, for the formation of which no concrete physical mechanism is known. However, should $\ell$ be stabilized by a longitudinal
$4$-dimensional physics mechanism\footnote{There exist both field and string theory 
arguments of this type~\cite{Antoniadis:2002tr,Dvali:2000xg,Kiritsis:2001bc}.},
then $\ell\sim M_{4}^{-1}$ and (up to factors of ${\cal O}(1)$)
\begin{equation}
   \L_{4}\sim
   M_{4}^{~4}~\Big(\frac{M_6}{ M_{4}}\Big)^{8}~.
   \label{e:LongCC}
\end{equation}
The original scenario of Ref.~\cite{rBHM5} then applies, where the
10-dimensional spacetime of the Type~IIB string theory is
compactified on a 4-dimensional supersymmetry preserving
space\footnote{All remaining
supersymmetry will be broken by the cosmic brane
solution~\cite{rBHM1}.} of characteristic size
$M_{10}^{-1}=M_6^{-1}\sim(10\,\mbox{TeV})^{-1}\sim10^{-19}$\,m. The
cosmic brane of Ref.~\cite{rBHM5} then describes a 3+1-dimensional
de~Sitter
spacetime, with the characteristic scale
$M_4\sim10^{19}\,$\,GeV. 
Furthermore, $L\define\Lb^{-1/2}\sim 10^{41}\,\mbox{GeV}^{-1}\sim 10^{25}$\,m, provides a natural scale which coincides with the Hubble radius.

Conversely, although the generic \BHM{} models break supersymmetry, their parameter space contains the $\w\,{\to}\,0$ supersymmetric and well understood limits~\cite{rFTh,rS-ForFld} and~\cite{Einhorn:2000ct} (see~\eqref{e:transII}, below), wherein
\begin{equation}
  \Big(\Lb \sim M_P^4\,(M_\text{susy}/M_P)^8\Big)~\to~0,\qquad
  \text{i.e.,}\qquad
  \Lb,M_\text{susy}\to0.
 \label{e:w->0}
\end{equation}
The \BHM{} deformation family of models are thus explicitly constructed as supersymmetry-breaking and positive (de~Sitter) cosmological constant inducing deformations of the familiar supersymmetric (and Minkowski) configurations in F-theory.

\paragraph{The Main Point:}
Given the above features of our solution, the main new point we want to make regarding the existence of de~Sitter backgrounds in string theory is as follows: String theory has purely stringy degrees of freedom (for example, the difference between the left and right string modes) not captured by the usual effective field theory/spacetime description used in the standard discussions, based on string compactifications, regarding the problem of de~Sitter space in string theory \cite{Kachru:2003aw, Balasubramanian:2005zx, Douglas:2006es}.
Such purely stringy (and, in general, non-commutative) degrees of freedom are captured in a phase-space formulation of string theory
~\cite{Freidel:2013zga, Freidel:2014qna, Freidel:2015pka, Freidel:2015uug, Freidel:2016pls, Freidel:2017xsi, Freidel:2017wst, 
Freidel:2017nhg, Freidel:2018apz}.
In that formulation, we can integrate over such dual/``momentum'' string degrees of freedom in order to generate an  effective spacetime description. Our proposal (for which we have some evidence from our solution, but not a detailed proof, as discussed in what follows) is that this procedure will naturally lead to an effective de~Sitter background, by inducing an effective dilaton (captured by the above $\w\q$-dependence) background which corresponds to the dilaton profile of our solution. This effective dilaton profile is anisotropic and the degree of anisotropy captures a positive cosmological constant of our solution.

The resulting de~Sitter background can be understood as a geometric deformation interpolating between clearly understood supersymmetric backgrounds of F-theory. (In addition, as we argue below, the resulting de~Sitter space can be also understood as a blow-up resolution of a singular Minkowski limit.) 
The above see-saw like relation between the cosmological constant scale and the scale of gravity and the scale of particle physics/supersymmetry breaking, is set by the requirements of a stringy $\SL(2;\ZZ)$ monodromy (as we shall argue in what follows)  and in particular the S-duality part of it, by viewing our solution as an S-fold\footnote{The notion of S-folds was originally coined by Hull in~\cite{rH-nGeoCY} being the S-duality analog of the T-folds~\cite{Dabholkar:2002sy,Flournoy:2004vn,
rHIS-nGeoCY}.} in the context of the recent phase-space
formulation of string theory~\cite{Freidel:2013zga, Freidel:2014qna, Freidel:2015pka, Freidel:2015uug, Freidel:2016pls, Freidel:2017xsi, Freidel:2017wst, Freidel:2017nhg, Freidel:2018apz} (and earlier, in the double field theory approach~\cite{Tseytlin:1990nb,Tseytlin:1990va,Siegel:1993th,Siegel:1993xq,Hull:2009mi}).
Note that this proposal goes around the usual picture of having an effective potential for some fields/moduli that produces vacuum expectation values for those fields (usually the vacuum being of a supersymmetric (AdS) kind and the de~Sitter solution being a long lived excitation around that vacuum produced by some stringy configurations, such as brane-antibrane systems, and stabilized by fluxes \cite{Kachru:2003aw, Balasubramanian:2005zx, Douglas:2006es}). In our proposal, we have, in principle, a cosmological de~Sitter
blow-up of a singular Minkowski background and a cosmologically induced supersymmetry breaking, and the hidden phase-space formulation operating behind the scenes and relating various energy scales via a natural see-saw like formula.

\section{Viability and Features}
\label{s:Features}
The \BHM\ deformation family of models described in Section~\ref{s:BHM} has some rather unusual features, which we now discuss in turn.

\paragraph{Viability:}
The \BHM{} models~\eqref{e:metric} evade the oft-quoted no-go theorem~\cite[Section~6]{Maldacena:2000mw} (see also Ref.~\cite[pp.~480--482]{rBBS} or the recent exhaustive review~\cite[Section~12.5]{Gran:2018ijr}) primarily by being non-compact: Following the stringy cosmic strings analysis~\cite{rGSVY,rCYCY}, we may compactify $\sY^2$ by including the limit-points $z\to\pm\infty$, and reexamine the behavior of the warp-factors over this now compact ``internal'' space.
 In particular, neither the $\Lb\,{>}\,0$ warp factors~\eqref{e:newAB} nor their $\Lb\,{\to}\,0$ variant~\eqref{e:oldAB} vanish at $z\to\pm\infty$ as would be required of these singularities in a compactification of $(\sY^2\cup\{\pm\infty\})\approx\IP^1$~\cite[Section~6.2]{Maldacena:2000mw}.
 In addition, Ref.~\cite[Section~6.2]{Maldacena:2000mw} emphasizes the importance of higher-curvature terms in~\eqref{e:Seff}---which are known to enable the evasion of the no-go theorem. It would be clearly desirable to determine their effect on the \BHM\ deformation family of models, but this remains an open question for now. Similarly, we defer the precise F-theory/heterotic dual of these considerations, and expected evasion of the dual no-go results~\cite{Green:2011cn,Kutasov:2015eba} to a subsequent effort\footnote{These standard no-go theorems follow within the context of supergravity
and the stringy $\alpha'$ corrections. We will give a general comment
about how the non-commutative phase-space formulation of string 
theory goes around this standard set-up at the end of Section~\ref{s:Other}.}.

 In turn, this family of models are driven by a highly nontrivial source: the energy-momentum tensor of the $\t\,{=}\,\t(\q)$ configurations~\eqref{e:moduli} is non-zero and forces the metric~\eqref{e:metric} to not be Ricci-flat~\eqref{e:EinStein}. In particular, the axion-dilaton configuration~\eqref{e:moduli} provides an {\em\/exotic matter\/} background, since its energy-momentum tensor is indefinite over $\sY^2$~\cite{rBHM2}:
\begin{equation}
 [\,T_{\mu\nu}\,]=[\,T(r)\,\eta_{ab}\,]\oplus
  \mathop{\mathrm{diag}}[\,-\w^2 r^{-2},~\w^2\,].
\label{e:exotic}
\end{equation}
This violates several of the energy positivity conditions, though not within $\sW^{3,1}$, reminding of the standard characteristics of traversable Lorentzian wormholes~\cite{rMV-LWh}.

\paragraph{Holomorphic Limits:}
As indicated by the deformation from step~\ref{i:Holo} to~\ref{i:nHolo} in the roadmap in Section~\ref{s:BHM}, the non-holomorphic axion-dilaton configurations~\eqref{e:moduli} are a deformation of the holomorphic configuration achieved by removing the anisotropy: $\w\,{\to}\,0$. This results in the familiar and well understood type IIB orientifold limit of F-theory~\cite{rFTh,rS-ForFld} with $\t=\a{+}i\,e^{-\F}=\textit{const}$.

However, starting with the configuration~\eqref{e:moduli}, it is also possible to find another novel holomorphic solution, by enforcing the Cauchy-Riemann conditions
\begin{equation}
  \pd{u}{r}=\frac1r\pd{v}{\q} \qquad\text{and}\qquad \pd{v}{r}=-\frac1r\pd{u}{\q},
  \qquad\text{for}~~\bigg\{
   \begin{array}{@{}r@{~}l}
    u(r,\q)&=\Re[f(r,\q)],\\[2pt]
    v(r,\q)&=\Im[f(r,\q)],
   \end{array}
 \label{e:CR}
\end{equation}
in systematic iterations. For example and definiteness, start with $\t_{I\!I}(\q)=-\t_0\tanh(\w\q)+i\,\t_0\sech(\w\q)$, i.e., with $u_0(r,\q)=\t_0\tanh(\w\q)$. The second of the Cauchy-Riemann conditions in~\eqref{e:CR} then implies that
\begin{alignat}9
   \pd{v}{r} &= \frac{\t_0}{r}\tanh'(\w\q),\qquad \tanh'(\w\q)\define\pd{}\q\tanh(\w\q),
  \nonumber\\
  \text{so}\quad
  v_1(r,\q)&=\t_0\log(r)\tanh'(\w\q)+f(\q),
 \label{e:v1}
\end{alignat}
where $f(\q)$ is an unknown integration $r$-constant. Solving in turn the first Cauchy-Riemann equation~\eqref{e:CR} for $u(r,\q)$, we complete the first iteration:
\begin{equation}
 u_1(r,\q)=\frc12\t_0\log^2(r)\tanh''(\w\q) +\log(r)f'(\q) +g(\q),
\end{equation}
where $g(\q)$ is another integration $r$-constant. This $u_1(r,\q)$ matches the supersymmetric axion~\cite{Einhorn:2000ct} in the limit $\w\to0$ upon choosing $\t_0\,{\mapsto}\,n/2\p\w$ and $g(\q)\,{\mapsto}\,u_0(r,\q)\,{=}\,\t_0\tanh(\w\q)$. With this $u_1(r,\q)$, we compute $v_2(r,\q)$ from the first and $u_2(r,\q)$ from the second Cauchy Riemann equation~\eqref{e:CR}, and so on. Expanding also in $\q$, re-summing and combining with the result of the same procedure starting from $v_0(r,\q)=\t_0\sech(\w\q)$ produces:
\begin{equation}
  \t_{I\!I}=\t_0\big(\tanh(\w\q)+i\sech(\w\q)\big)\quad\to\quad
  \t_0\big(\tanh[\w(\q{-}iz)]+i\sech[\w(\q{-}iz)]\big),
 \label{e:D7def}
\end{equation}
with $z\,{=}\,\log(r)$ (and $\ell\,{=}\,1$ for simplicity).
 The non-holomorphic configuration~\eqref{e:tII} then has two distinct holomorphic limits, and interpolates between them:
\begin{equation}
 \mkern-12mu
  \begin{array}{@{}c@{}c@{}c@{}}
 \Big[b_0{+}\t_0\big(\tanh(\w\q){+}i\sech(\w\q)\big)\Big]
 &\xrightarrow[\q\,\to\,\q-iz]{\text{iterate~\eqref{e:CR}}}
 &\Big[b_0{+}\t_0\big(\tanh[\w(\q{-}iz)]{+}i\sech[\w(\q{-}iz)]\big)\Big]
 \\*[-3pt]
 \big\downarrow\rlap{$\vcenter{\hbox{$\scriptstyle\w\,\to\,0$}}$} &&
  \big\downarrow\rlap{$\vcenter{\hbox{$\scriptstyle O(\w)$}}$}\\*[2pt]
 \Big[(\a,e^{-\F})=(b_0,\t_0)\Big]_{\!\!\text{\scriptsize\cite{rFTh,rS-ForFld}}}
 &\leftarrow\joinrel\xrightarrow[\smash{\mkern-12mu\eqref{e:tII}:~\t_{I\!I}}]
                                 {\mkern-12mu\text{transition~}}&
   \Big[\big(\, [b_0{+}\t_0\w\q] \,,\, \t_0[1{-}\w z] \,\big)\Big]_{\!\!\text{\scriptsize\cite{Einhorn:2000ct}}}.
  \end{array}\mkern-12mu
 \label{e:transII}
\end{equation}
This then defines the \BHM\ $\t_{I\!I}$-transition, which interpolates between the constant axion-dilaton configurations~\cite{rFTh,rS-ForFld} and the (``helicoidal axion'') D7 instanton~\cite{Einhorn:2000ct} after choosing $\t_0\mapsto\frac{n}{2\p}g_s^{-1}$ and identifying $\w\mapsto g_s$.
 This also relates~\eqref{e:tII} to its ``holomorphization''~(\ref{e:transII}, top-right), implemented by the simple analytic continuation $\q\to\q{-}i\log(r/\ell)$, which then itself provides a (third) related holomorphic Ansatz for the axion-dilaton system, and so a candidate related supersymmetric configuration.

 The same can be done with the other \BHM\ solution~\eqref{e:tI}, also resulting in the straightforward analytic continuation $\t_I(\q)\to\t_I(\q{-}iz)$. The resulting analogous \BHM\ $\t_I$-transition is:
\begin{equation}
  \begin{array}{c@{~~}c@{~~}c}
 \Big[b_0{+}i\t_0e^{\w\q}\Big]
 &\xrightarrow[\q\,\to\,\q-iz]{\text{iterate~\eqref{e:CR}}}
 &\Big[b_0{+}i\t_0e^{\w(\q-iz)}\Big] \\*[-3pt]
 \big\downarrow\rlap{$\vcenter{\hbox{$\scriptstyle\w\,\to\,0$}}$} &&
  \big\downarrow\rlap{$\vcenter{\hbox{$\scriptstyle O(\w)$}}$}\\*[2pt]
 \Big[(\a,e^{-\F})=(b_0,\t_0)\Big]
 &\leftarrow\joinrel\xrightarrow[\smash{\mkern-12mu\eqref{e:tI}:~\t_I}]
                                 {\mkern-12mu\text{transition~}}&
   \big(\, [b_0{+}\t_0\w z] \,,\, \t_0[1{+}\w\q] \,\big),
  \end{array}
 \label{e:transI}
\end{equation}
wherein the bottom-right corner, $O(\w)$-configuration has a $\q$ vs.\ $z{\define}\log(r/\ell)$ dependance that is $(z,\q)\to(-\q,z)$ {\em\/rotated\/} from the one in~(\ref{e:transII}, bottom-left).

\paragraph{Stability,~1:}
For the axion-dilaton configurations~\eqref{e:moduli} to specify stringy rather than merely supergravity solutions, their parameters $b_0,\w,g_s$ 
must be restricted so that $\t=\t(\q)$ would exhibit an $\SL(2;\ZZ)$-monodromy rather than a continuous $\SL(2;\IR)$-transformation.
Thus furnishing discrete $\SL(2;\ZZ)$-orbits and since $\dim\SL(2)\,{=}\,3$, the 3-parameter family of choices~\eqref{e:moduli} $\t=\t(\q;b_0,\w,g_s)$ is naturally expected to form a {\em\/discretuum\/}. 
 Since the axion-dilaton system naturally couples to fluxes, the well-known eponymous string theory results~\cite{rP+S-Discretuum,rBP-Discretuum,Giddings:2001yu}  corroborate the discreteness of the \BHM\ configurations~\eqref{e:moduli}, as do the general string theory expectations~\cite{rBS-Gauge}; however, we are not aware of a rigorous proof. This implies that most of the continuous $(b_0,\w,g_s)$-parameter space is the non-stringy ``swampland''---except for the discrete subset of $\SL(2;\ZZ)$-isolated points within it.

 With this in mind, the \BHM\ family of models can vary continuously only via the metric parameters $z_0$ and $\ell$ (i.e., $\Lb$), which are independent of $b_0,\w,g_s$.
 Eqs.~\eqref{e:newAB}, \eqref{e:link} and the $\sY^2$-geometry that they parametrize (see Figure~\ref{f:charts}) and the fact that $z_0$ is at infinite proper distance~\cite{rBHM2} jointly imply that the transversal length-scale $\ell$ in $\sY^2$ remains the only continuously variable physically relevant parameter.
 As discussed above, in justifying~\eqref{e:LongCC}, no definitive physical mechanism is known for stabilizing $\ell$, although there do exist arguments to this end both in field and in string theory~\cite{Antoniadis:2002tr,Dvali:2000xg,Kiritsis:2001bc}; see also below.
 All this provides the \BHM\ models with an unexpected and high degree of stability.

\paragraph{Anisotropy:}
The $\SL(2;\ZZ)$ monodromy requires that $(g^D_s\define\vev{e^{-\F}}_{\sW\rtimes\sY^2})\sim O(1)$, which also agrees with modular invariance: the \BHM\ models require string theory to not be weakly coupled throughout $\sW\,{\rtimes}\,\sY^2$. 
We therefor expect higher order corrections.
 Nevertheless, within the $(D{-}2)$-dimensional spacetime
\begin{equation}
  @~\sW^{D-1,1}_{\!z=0}:~~g_s^{D-2}=g_s^D\sqrt{\a'/V_\perp}\ll1,\quad V_\perp\define\mathrm{Vol}(\sY^2),
 \label{e:weakST}
\end{equation}
since $V_\perp$ may be chosen to be large~\cite{rBHM1}: String theory is thus weakly coupled within $\sW^{D-1,1}_{\!z=0}$, and therein the low-energy effective field theory approximation is well justified.

In going beyond the tree-level approximation
consider evaluating the string theory scattering cross-section for any particular process in the $D$-dimensional spacetime, where $g_s^D\sim O(1)$. In this double expansion, ordered by powers of $\a'$ and of $g_s$, the latter is equivalently ordered by the genus of the interacting worldsheet surface. Compare now the genus-$g$ contributions in any such computation with those at genus-$(g{+}1)$ ---{\em\/with everything else the same}. In a straightforwardly pragmatic sense, the relative ratio of such two contributions provides a measure as to how reliable string-perturbative computations are, i.e., how weakly (or strongly) string theory is coupled; dub this the {\em\/effective\/} string coupling parameter $g_s^{\sss\text{(eff)}}$.

 All such ratios (for any particular physical process) will necessarily depend on the {\em\/local\/} value of the dilaton field\footnote{The ``running'' of coupling parameters, i.e., the dependence of the interaction strength on the colliding momenta is of course familiar in quantum field theory, and naturally translates also into a dual dependence on the collision proximity. While different in technical details, the fact that the dilaton and the string interaction strength can vary over the position (and momentum space) is conceptually the same.} (and possibly also the axion). In the \BHM{} models, these fields vary over the $D$-dimensional spacetime, and so does then also this {\em\/effective\/} string coupling, $g_s^{\sss\text{(eff)}}$. Notably, the \BHM{} configurations~\eqref{e:moduli} imply $g_s^{\sss\text{(eff)}}\,{<}\,1$ in some $\q$-directions in $\sY^2$, $g_s^{\sss\text{(eff)}}\,{\sim}\,1$ in others, and even $g_s^{\sss\text{(eff)}}\,{>}\,1$ within the $\t_I$ configuration:
\begin{equation}
   g_s^{\sss\text{(eff)}}[\t_I(\p{-}\e)]\ll1
    \qquad\text{whereas}\qquad
   g_s^{\sss\text{(eff)}}[\t_I(\p{+}\e)]\gg1.
 \label{e:gsJumps}
\end{equation}
The \BHM{} models thereby explicitly patch effectively weakly-coupled string theory to effectively (S-dual) strongly-coupled string theory across the $\q=\p$ direction in $\sY^2$.\footnote{This is akin to the T-fold solutions~\cite{Dabholkar:2002sy,Flournoy:2004vn,rH-nGeoCY,rHIS-nGeoCY}, where local chart patching involves (T-duality) mirror-symmetry.}
With $g_s^D\sim O(1)$ and by gluing regimes with reciprocally weak/strong effective string interactions~\eqref{e:gsJumps}, this S-duality patching makes it evident that the low-energy tree-level effective field theory approximation is sorely lacking. 

Given the important role of modular invariance and the $\SL(2;\ZZ)$ 
transformations for our solution, and in particular, the S-duality part
of $\SL(2;\ZZ)$, we want to draw an analogy with what is known about T-duality 
and T-folds in the context of the double field theory, and the phase-space
formulation of string theory. We propose that our solution should be 
naturally viewed from a phase-space point of view, as an S-duality analog
of the T-fold, that we call an S-fold, which glues the weakly and the strongly coupled regimes of our solution\footnote{The use of S-folds has also recently occurred in the context of SCFT in various dimensions, e.g.,~\cite{Aharony:2016kai}.}. In order to make sense of this
picture, we need to include stringy degrees of freedom required for 
such a phase-space formulation that are not taken into
account in the effective field theory discussion
In particular, the usual identification of the target space being spanned by only the sum
 $\vev{\hat{X}^\mu_L(\t,\s){+}\hat{X}^\mu_R(\t,\s)}$ should be amended.
The S-dual (and strongly stringy-coupled) patching~\eqref{e:gsJumps} indicates a need for (re)incorporating other stringy degrees of freedom, and at the very least also
 $\vev{\hat{X}^\mu_L(\t,\s){-}\hat{X}^\mu_R(\t,\s)}$:
These vev's being determined by linear combinations of the canonically conjugate/dual Schr{\"o}\-din\-ger center-of-mass operators~\cite{rJPS} $\hat{x}^\mu$ and $\hat{p}^\mu(\t/p^+)$ implies the need to (re)incorporate the ``momentum'' space into this more complete description of the target space in string theory; we return to this in the next section. This then leads to a (non-commutative) phase-space geometry of the type discussed in Refs.~\cite{Freidel:2013zga, Freidel:2014qna, Freidel:2015pka, Freidel:2015uug, Freidel:2016pls, Freidel:2017xsi, Freidel:2017wst, Freidel:2017nhg, Freidel:2018apz}.

\section{Generalizations and Implications}
\label{s:Other}
\paragraph{Diversity:}
As indicated at the outset~\cite{rBHM1}, analogous models can be built driven by other moduli fields, $\f^\a$.
 This generalization then allows the $\sY^4$-moduli to (also) vary over $\sY^2=S^1{\times}\sZ$.
 If $\f^\a=\f^\a(\q)$ is again aperiodic and anisotropic, and since the Weil-Petersson-Zamolod\-chikov metric in moduli spaces of Calabi-Yau varieties is the natural generalization of the Teichm{\"u}ller metric~\cite{rCHS-ZWP}, $\f^\a(\q)$ will have to exhibit a ``mapping class group'' monodromy, generalizing $\SL(2;\ZZ)$; see, e.g., Ref.~\cite{rMirr00,rMirr01} for a concrete example.
 We then expect the dynamics and phenomenology to be similar to the one driven by $\t=\t(\q)$.
 It of course remains to verify that $\ell\sim\text{size}(\sZ)$ may be chosen so as to satisfy the experimental limits on extra dimensions while preserving other desirable phenomenological features.

 For example, choosing $\f=\f(\q)$ to be the size (``breathing'') modulus of $\sY^4$, the model patches, akin to~\eqref{e:gsJumps}, the ``large'' and the ``small'' copies of $\sY^4$ across the identified endpoints of $\q\in[-\p,\p]$. This is precisely the crux of the T-fold constructions~\cite{Dabholkar:2002sy,Flournoy:2004vn,rH-nGeoCY,rHIS-nGeoCY}. Conversely then, it is natural to ask whether these already generalized constructions can be deformed akin to the \BHM\ models.
 On the other hand, the corresponding analogue of the ``holomorphization''~\eqref{e:D7def}, as implemented in~\eqref{e:transII} by the analytic continuation $\q\to\q{-}i\log(r/\ell)$ eerily reminds of the analytic continuation $J\to J{+}iB$ of the K{\"a}hler form, which has ever since~\cite{rAGM00,rAGM03,rAGM04} and especially~\cite{rPhases,rMP0} become {\em\/sine qua non,\/} in the study of moduli spaces of Calabi-Yau $n$-folds. 
 
Conversely then, the non-holomorphic and anisotropic configurations~\eqref{e:moduli} driving the \BHM\ deformation family of models~\eqref{e:metric}+\eqref{e:newAB} are easily seen to be analogous to the (exceptional) $B\to0$ limit
in the by now much better understood moduli space of Calabi-Yau $n$-folds. The fact that it is this non-holomorphic and anisotropic configuration~\eqref{e:moduli} that also parametrizes both supersymmetry breaking and the possibility of the Minkowski\,$\to$\,de~Sitter deformation warrants a closer analysis of these \BHM\ models.

\paragraph{Genericity:}
The Euclidean analytic continuation of the 9+1-dimensional spacetime in the type IIB string theory
has to be Ricci-flat, whether compact or not, and hence is a (possibly non-compact) Calabi-Yau 5-fold.
This certainly resonates with the geometric quantization conclusion, that Ricci-flatness of the loop-space is ``the string equation of motion''~\cite{rFrGaZu86,rBowRaj87,rBowRaj87a,rBowRaj87b,rHHRR-sDiffS1,Pilch:1987eb}.

Just as Calabi-Yau 3-folds generically contain many isolated $\IP^1$'s, the so-called $\mathcal{O}(-1,-1)$-curves, which are {\em\/small resolutions\/} of {\em\/nodes,\/} (i.e., double-points, $A_1$-singularities, conifold singularities)~\cite{rBeast}, Calabi-Yau 5-folds generically contain many isolated Fano ($c_1>0$) compact complex surfaces $\cS$~\cite{rHitch}. For a Calabi-Yau 3-fold, these  exceptional $\mathcal{O}(-1,-1)$-curves have a bulk K{\"a}hler metric which is generally null, but is straightforwardly deformed into a positive metric by adding a multiple of the intrinsic volume form~\cite{rGHSAR}. Analogously, the bulk K{\"a}hler metric of the above Calabi-Yau 5-fold is null on the exceptional complex surfaces $\cS$, but is straightforwardly deformed into a positive metric by adding a multiple of the the K{\"a}hler metric specified by the intrinsic volume form of $\cS$, which we dub the ``bulk+local metric deformation.''

Analytically continuing this Euclideanized Ricci-flat 10-fold back to a 9+1-dimensional spacetime, at least some of the generically occurring exceptional complex surfaces $\cS$ within Calabi-Yau 5-fold will map to 3+1-dimensional sub-spacetimes.
 Within these 3+1-dimen\-si\-onal spacetime bubbles, the ``bulk+local metric deformation'' would naturally correspond  to the Minkow\-ski\,$\to$\,de~Sitter desingularization deformation discussed in Section~\ref{s:BHM}, with $\Lb>0$ parametrizing the size of the analytically continued $\cS$. This is indeed the deformation employed within the uncompactified four dimensional spacetime $\sW^{3,1}$ in Section~\ref{s:BHM}, as discussed right after~\eqref{e:oldAB}. 
We conjecture that a very similar, ``de~Sitter-izing,'' supersymmetry breaking metric deformation may be employed within at least some of the {\em\/generically plentiful\/} exceptional surfaces $\cS$.

\paragraph{Seeing Double:}
In any worldsheet field theory underlying string theory, the large modes (with wavelengths $\lambda>\ell_s\mathop{:=}\sqrt{\alpha'}$) of both coordinate fields $\hat{X}^\mu_L$, $\hat{X}^\mu_R$ probe the
target space $\sX$, whereas stringy-small modes (with $\lambda<\ell_s\mathop{:=}\sqrt{\alpha'}$) of both $\hat{X}^\mu_L$, $\hat{X}^\mu_R$ probe $\widetilde{\sX}$, the mirror spacetime. 
The full target space of string theory is therefore (locally) a product of these two factors, the latter of which is naturally identified to be the (T-dual) mirror of the former.
 This strongly resonates with the phase-space theory discussed at the end of Section~\ref{s:Features} and in Refs.~\cite{Freidel:2013zga, Freidel:2014qna, Freidel:2015pka, Freidel:2015uug, Freidel:2016pls, Freidel:2017xsi, Freidel:2017wst, Freidel:2017nhg, Freidel:2018apz}, which may be justified conceptually also as follows: 
\begin{enumerate}\itemsep=-3pt\vspace*{-1mm}
 \item Consider a point-particle moving on a rigid circle.
  \begin{enumerate}\itemsep=-3pt\vspace*{-1mm}
   \item At first blush, the phase-space of a point-particle on a circle is a cylinder, where the (vertical) $\IR^1$-like $p_\f$-generator represents momentum, at each point of the circle of possible positions, $\f$.
   \item However, $p_\f\to+\infty$ and $p_\f\to-\infty$ are indistinguishable: if one moves infinitely fast, it does not matter in which direction one is moving. This compactifies the momentum direction into a circle, and the phase-space into a ring-torus (adding a copy of the position-circle at $p_\f$-infinity).
   \item However, when one moves infinitely fast, one is simultaneously everywhere, so that the positional circle at $p_\f$-infinity shrinks to a point.
\vspace{-5pt}
\end{enumerate}
The final, (b)\,$\to$\,(c) step in this progression of modeling the phase-space of a point-particle on a circle thus looks as in Figure~\ref{f:Tori}.
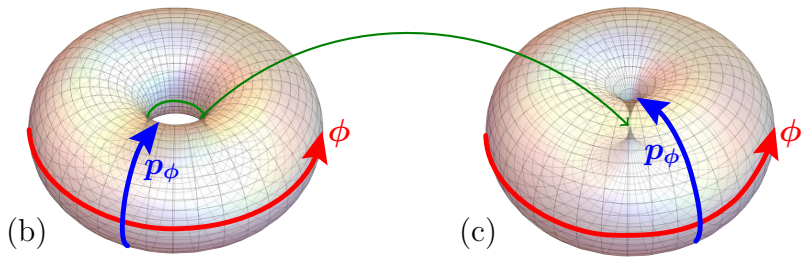
\begin{figure}
 \begin{center}
  \TikZ{\path[use as bounding box](-5,-1.25)--(5,1.75);
            \path(-3,0)node{\includegraphics[height=35mm]{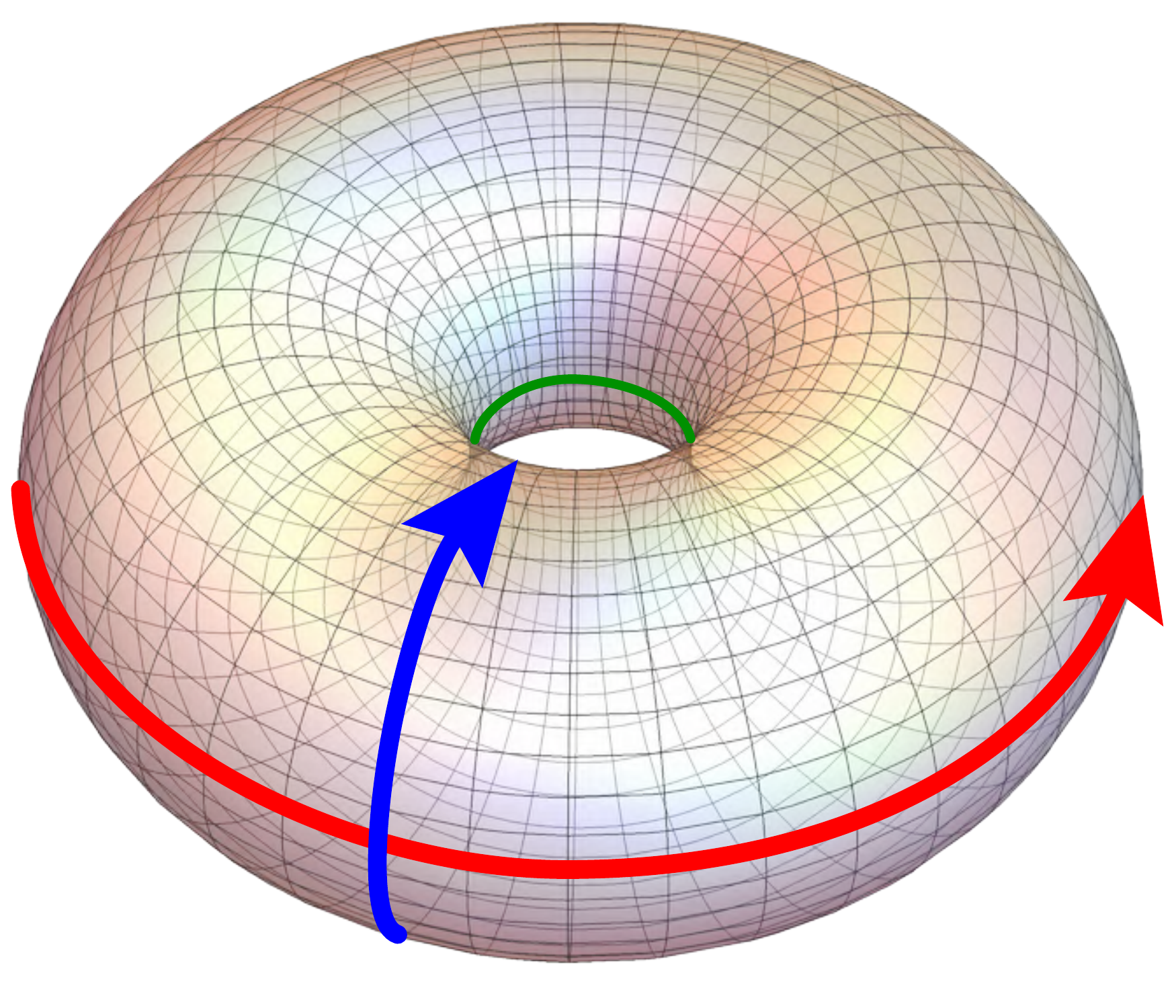}};
            \path(3,0)node{\includegraphics[height=35mm]{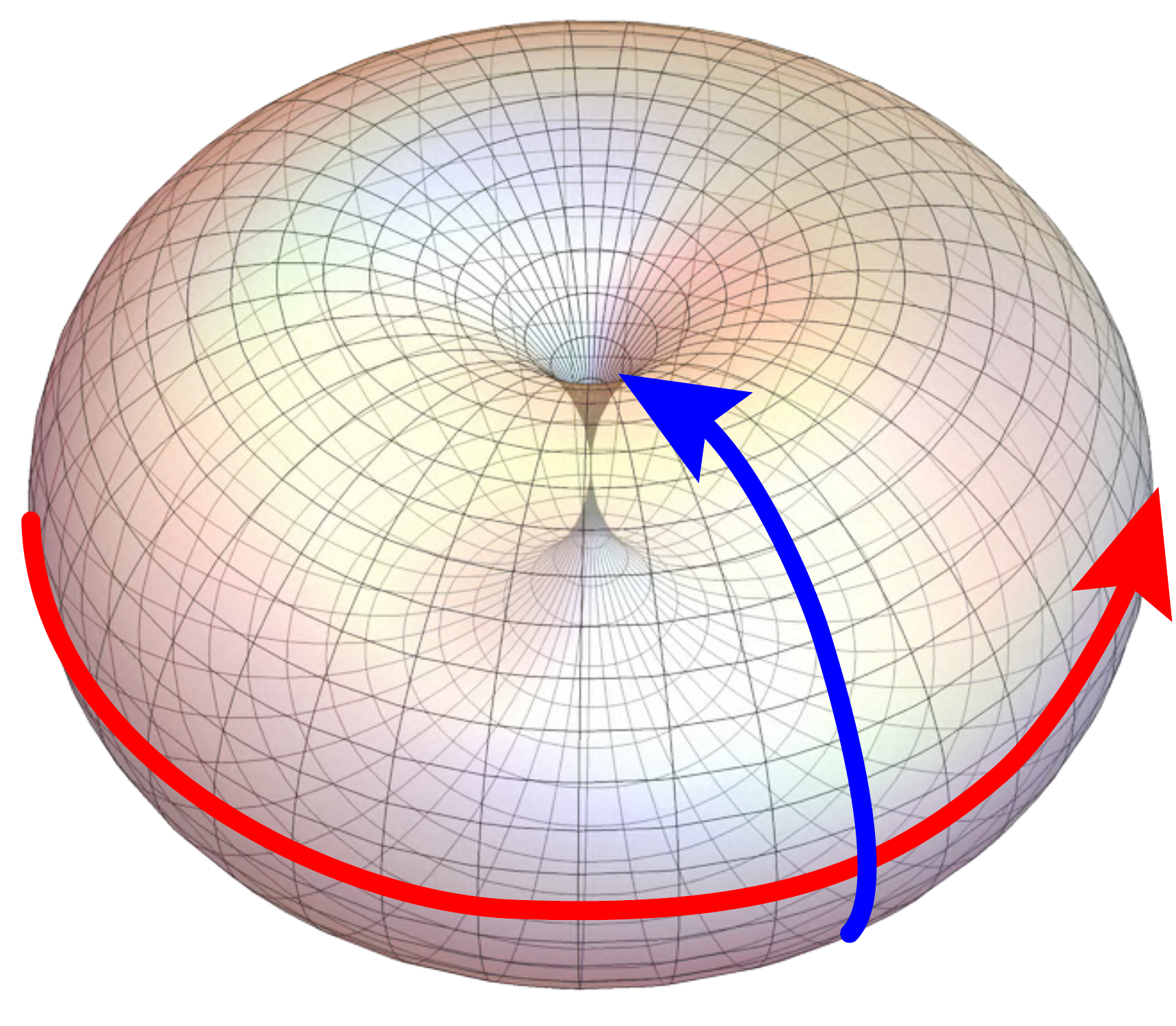}};
            \path[red](-.85,0)node{$\boldsymbol{\f}$};
            \path[blue](-3.2,-.5)node{$\boldsymbol{p_\f}$};
            \path[red](5.15,0)node{$\boldsymbol{\f}$};
            \path[blue](3.45,-.3)node{$\boldsymbol{p_\f}$};
            \draw[thick,green!50!black,->](-2.7,.2)to[out=45,in=135]++(5.7,-.1);
            \path(-5,-1.33)node{(b)};
            \path(1,-1.33)node{(c)};
           }
 \end{center}
\caption{The phase-space of a point particle moving along a circle with fixed radius, at finite speed in b), and at infinite speed in c).}
 \label{f:Tori}
 \end{figure}

 \item The configuration space of a particle moving on an $n$-torus is $T^n=\IR^n/\L$, whereas its momentum space is naturally the (mirror) dual torus, $\widetilde{T}{}^n=\IR^n/\L^*$.
 This makes the phase-space into $T^{2n}\sim\IR^{2n}/(\L\oplus\L^*)$, ---except that again the total space is singular ``at infinity.''

 \item For Calabi-Yau varieties, there exists a conjecture\footnote{Dec.~1993, Lexington: \url{http://www.ams.org/meetings/sectional/1890_program_ssh.html\#title}}:
  \begin{enumerate}\itemsep=-3pt\vspace*{-1mm}
   \item a completion of each fiber of $T^*_{\!M}$ is a mirror of the Calabi-Yau variety $M$;
   \item a completion of $T^*_{\!M}$ is a ``double Calabi-Yau space,'' singular ``at infinity.''
  \end{enumerate}
\end{enumerate}

This would imply the ``true home'' for the above-cited phase-space $(x,\tilde{x})$-geometry to be such a singular {\em\/foliation\/}, $\sX{\divideontimes}\widetilde{\sX}$, of the usual spacetime $\sX$ and its mirror $\widetilde{\sX}$: Locally at any generic point, $\sX{\divideontimes}\widetilde{\sX}$ looks like a product, but over certain locations in one factor, the other may singularize. By mirror symmetry, neither of the two factors is a preferred base of a fibration, and the total space may well be singular---as is the right-hand side illustration in Figure~\ref{f:Tori}. Furthermore, $\sX{\divideontimes}\widetilde{\sX}$ is naturally endowed with a non-commutativity structure induced ultimately from the symplectic position-momentum structure on the worldsheet, as employed in Refs.~\cite{Freidel:2013zga, Freidel:2014qna, Freidel:2015pka, Freidel:2015uug, Freidel:2016pls, Freidel:2017xsi, Freidel:2017wst, Freidel:2017nhg, Freidel:2018apz}.
 Incidentally, this target space doubling is also suggested from a closer look at geometric quantization, since it in fact involves the {\em\/oriented\/} (and so doubled) loop-space~\cite{rFrGaZu86,rBowRaj87,rBowRaj87a,rBowRaj87b,rHHRR-sDiffS1,Pilch:1987eb}.
 Clearly, the $\IR^{3,1}$-factor and its dual in $\sX{\divideontimes}\widetilde{\sX}$ are diffeomorphic, but may well have different ---and presumably complementary--- metric properties, the study of which we defer for now.

\paragraph{Stability,~2:}
To recap, our \BHM\ models that realize a positive cosmological constant within string theory depend on two types of parameters: The $b_0,\w,g_s$ parametrizing the axion-dilaton system~\eqref{e:moduli} are restricted by modular invariance to a discrete subset of $\SL(2;\ZZ)$-isolated points.
 In turn, the metric~\eqref{e:metric}--\eqref{e:newAB} depends on $\ell$, which can be stabilized by providing a phase-space interpretation of our model, via the requirement of T-duality (which is fully covariant in the phase space/non-commutative formulation of string theory~\cite{Freidel:2013zga, Freidel:2014qna, Freidel:2015pka, Freidel:2015uug, Freidel:2016pls, Freidel:2017xsi, Freidel:2017wst, Freidel:2017nhg, Freidel:2018apz}. 
Both of these features are intrinsically stringy.

Second, the see-saw formula~\eqref{e:LongCC} for the cosmological constant in our model can be also understood as an example of a formula required by T-duality, given, first, the scale of non-commutativity of the phase space formulation~\cite{Freidel:2013zga, Freidel:2014qna, Freidel:2015pka, Freidel:2015uug, Freidel:2016pls, Freidel:2017xsi, Freidel:2017wst, Freidel:2017nhg, Freidel:2018apz} (set by an effective size of the string), and captured by the parameter $\w$ in our model, and, second, given the Planck scale (set by the value of the dilaton, viewed as the relevant volume form of the stringy phase space), and captured by the parameter $\ell$ of our model.

More intuitively, the stringy de~Sitter space can be understood as a blow-up in the ``Calabi-Yau-ization'' of the 10 dimensional Minkowski habitat of string theory, as implied by the ``Einstein equation'' in the stringy loop space, which implies an infinite dimensional Ricci-flat/Calabi-Yau nature of the stingy configuration space~\cite{rFrGaZu86,rBowRaj87,rBowRaj87a,rBowRaj87b,rHHRR-sDiffS1,Pilch:1987eb}. 
The value of the cosmological constant is then the size of the relevant blow-up fixed by the requirements of T-duality (``mirror symmetry'') in the phase space reformulation of 
the geometry of the stringy loop space. Then the basic idea is that the stringy stability  of a ``stringy de~Sitter spacetime'' follows from an optimization between the short- and the long-distance spacetime physics emerging from the phase-space formulation of string theory, the geometry of which is really responsible for the appearance of a positive cosmological constant in the first place.

The actual construction reviewed in this paper can be therefore understood as an illustrative toy model for this new, intrinsically non-commutative phase-space picture of string theory that naturally leads to de~Sitter backgrounds, and which we hope to explore in more detail in the sequel to this note. 

\section{Summary, Outlook and Conclusions}
\label{s:Coda}
In this reexamination of the \BHM\ deformation family of string compactifications,
we have summarized the salient features of this class of models as originally developed~\cite{rBHM1,rBHM2,rBHM3,rBHM4,rBHM5,rBHM6}, as well as teased out some previously unpublicized characteristics. 

In particular,
 ({\small\bf1})~the driving sources~\eqref{e:moduli} admit a ``holomorphization''~\eqref{e:D7def}, and thereby two separate supersymmetric limits~\eqref{e:transII}, whereby the \BHM\ configurations~\eqref{e:moduli} {\em\/interpolate\/} between these two distinct and well known supersymmetric configurations, as well as the third holomorphic and possibly supersymmetric configuration~\eqref{e:D7def}.
 In addition,
 ({\small\bf2})~the aperiodic anisotropy of the axion-dilaton configuration~\eqref{e:moduli} exhibits a type of chart-patching~\eqref{e:gsJumps} that explicitly employs S-duality, and so implies that the \BHM\ models cannot be limited to weak string coupling. It is important to note that
 ({\small\bf3})~the \BHM\ configurations~\eqref{e:moduli} may equally well be used for other moduli, where the aperiodic anisotropy implies chart-patching~\eqref{e:gsJumps} that employs T-duality---strikingly similar to the T-folds~\cite{Dabholkar:2002sy,Flournoy:2004vn,rH-nGeoCY,rHIS-nGeoCY}.
Also, the
 ({\small\bf4})~``holomorphization''~\eqref{e:D7def} bears a strikingly similarity to the by now very well understoof $J\,{\to}\,J{+}iB$ analytic continuation of K{\"a}hler moduli spaces of Calabi-Yau $n$-folds. By converse, the physics driven by the \BHM\ configurations are then fairly ubiquitous, and correspond to the $B\to0$ limit in the moduli space of Calabi-Yau $n$-folds. 
 Finally, the
 ({\small\bf5})~Euclidean version of the Minkowski\,$\to$\,de~Sitter deformation within the 
 3+1 dimensional spacetime
 $\sW^{3,1}_{\!z=0}$ seems strikingly similar to the deformation of the K{\"a}hler metric in the blowup or small resolution exceptional sets. Since exceptional complex 2-folds are ubiquitous (as small resolutions of nodes) within Calabi-Yau (complex) 5-folds, in a converse Lorentzian analytic continuation, at least some of those complex 2-folds could serve as 3+1-dimensional (sub)spacetimes---and should admit a de~Sitter metric \`a la~\eqref{e:metric}+\eqref{e:newAB}.

As explained in the main body of the paper, the \BHM\ solutions discussed herein can be viewed as a well-defined deformation of the stringy cosmic strings/branes in type IIB/F-theory, and the latter can, at least in principle, be related to a deconstruction of the cosmological constant from 3- to 4-dimensional spacetime. 
What is meant here is the old observation of Witten~\cite{Witten:1994cga} about the peculiar features of supersymmetry in 3-dimensional spacetime, where due to the presence of conical defects in 3-dimensional gravity, the supercharges do not have to be globally defined, so one has supersymmetry but  not the degeneracy in masses between bosons and fermions ---the mass splitting is controlled by the strength of the conical defect. By deconstruction (performed in~\cite{Jejjala:2002we}, but also discussed in~\cite{Becker:1995sp}), one can obtain a 4-dimensional version of Witten's argument, albeit now with stringy defects. These stringy defects are generically strongly coupled in the 4-dimensional continuum limit, and could be naturally related to the non-supersymmetric stringy cosmic strings.
 So, in principle, our construction can be connected to this narrative, even though the details remain to be worked out.

Finally, as discussed at the end of Section~\ref{s:Other}
 and motivating a sequel to this note,
 the latter few of the above observations imply that the familiar point-field limit description of spacetime is in fact incomplete: more of the stringy degrees of freedom must be included, extending the stringy target space into a certain double, tentatively modeled on the corresponding phase-space.

\paragraph{Acknowledgments:} 
We thank Lara Anderson and James Gray for interesting comments on string compactifications. DM thanks Laurent Freidel and Rob Leigh for numerous insightful discussions over many years on the topic of
quantum gravity and string theory, and Vijay Balasubramanian, Petr Ho{\v{r}}ava and Jan de Boer for many conversations on de~Sitter spacetime and string theory.
 PB would like to thank the CERN Theory Group for their hospitality over the past several years.
 TH is grateful to the Department of Physics, University of Maryland, College Park MD, and the Physics Department of the Faculty of Natural Sciences of the University of Novi Sad, Serbia, for the recurring hospitality and resources.
 The work of DM is supported in part by the Julian Schwinger Foundation.

\small
\raggedright

\end{document}